\documentclass[aps,prl,twocolumn,showpacs,floatfix,superscriptaddress]{revtex4}
\usepackage{graphicx}

\begin{document}
\flushbottom

\title{Phonon-mediated anisotropic superconductivity in the Y and Lu nickel borocarbides}
\author{P. Mart\'inez-Samper}
\affiliation{Laboratorio de Bajas Temperaturas, Departamento de F\'isica de
la Materia Condensada \\ Instituto de Ciencia de Materiales
Nicol\'as Cabrera, Facultad de Ciencias \\ Universidad
Aut\'onoma de Madrid, 28049 Madrid, Spain}
\author{H. Suderow}
\affiliation{Laboratorio de Bajas Temperaturas, Departamento de F\'isica de
la Materia Condensada \\ Instituto de Ciencia de Materiales
Nicol\'as Cabrera, Facultad de Ciencias \\ Universidad
Aut\'onoma de Madrid, 28049 Madrid, Spain}
\author{S. Vieira}
\affiliation{Laboratorio de Bajas Temperaturas, Departamento de F\'isica de
la Materia Condensada \\ Instituto de Ciencia de Materiales
Nicol\'as Cabrera, Facultad de Ciencias \\ Universidad
Aut\'onoma de Madrid, 28049 Madrid, Spain}
\author{J. P. Brison}
\affiliation{Centre des Recherches sur les Tr\`es Basses Temp\'eratures \\
CNRS, BP 166, 38042 Grenoble Cedex 9, France}
\author{N. Luchier}
\affiliation{Centre des Recherches sur les Tr\`es Basses Temp\'eratures \\
CNRS, BP 166, 38042 Grenoble Cedex 9, France}
\author{P. Lejay}
\affiliation{Centre des Recherches sur les Tr\`es Basses Temp\'eratures \\
CNRS, BP 166, 38042 Grenoble Cedex 9, France}
\author{P. C. Canfield}
\affiliation{Ames Laboratory and Departament of Physics and Astronomy \\
Iowa State University, Ames, Iowa 50011, USA}
\date{\today}

\begin{abstract}

We present scanning tunneling spectroscopy and microscopy
measurements at low temperatures in the borocarbide materials
RNi$_2$B$_2$C (R=Y, Lu). The characteristic strong coupling
structure due to the pairing interaction is unambiguously resolved
in the superconducting density of states. It is located at the
superconducting gap plus the energy corresponding to a phonon mode
identified in previous neutron scattering experiments. These
measurements also show that this mode is coupled to the electrons
through a highly anisotropic electron-phonon interaction
originated by a nesting feature of the Fermi surface. Our experiments, 
from which we can extract a large
electron-phonon coupling parameter $\lambda$ (between 0.5 and
0.8), demonstrate that this anisotropic electron-phonon coupling
has an essential contribution to the pairing interaction. The
tunneling spectra show an anisotropic {\it s}-wave superconducting
gap function.
\end{abstract}

\pacs{74.70.Dd, 74.62.Dh, 74.20.Mn}
\maketitle

In most known superconductors the formation of Cooper pairs is
based on an attractive interaction mediated by phonons. However,
Cooper pairing driven by other bosonic excitations has also
attracted much attention due to its fundamental interest. Some of
the proposed mechanisms have been used to interpret the
superconducting behavior of different compounds discovered during
the last decades. But clear microscopic information is very much
needed in this field. The situation is especially puzzling in the
borocarbide materials (RNi$_2$B$_2$C, R=Y, Lu, Tm, Er, Ho, Dy),
where recent experiments have shown that the question about the
pairing interaction, initially thought to be conventional
electron-phonon coupling\cite{Mattheiss94, Yanson97, Cheon99}, is
far from being understood. These compounds show moderate critical
temperatures (between 6K and 16.5K\cite{Cava, Canfieldgeneral}) and
very interesting phase diagrams where superconductivity coexists
with antiferromagnetic order (when R is a magnetic rare earth,
R=Tm, Er, Ho, Dy). The behaviors observed in the thermal
conductivity\cite{Boaknin01, Izawa02}, photoemission
spectroscopy\cite{Yokoya00}, specific heat\cite{Izawa01, Nohara00},
microwave surface impedance\cite{Izawa01, Jacobs} and
Raman-scattering\cite{SangYang} experiments of the non-magnetic Y
and Lu borocarbides (which also present the highest critical
temperatures of 15.5 K and 16.5 K respectively) show that the
superconducting gap is highly anisotropic. Indirectly, this could
be related to an also anisotropic pairing interaction, but no
experiment has given an indication of its nature.

On the other hand, Fermi surface nesting seems to be a general
feature of the whole family of borocarbide materials. It produces
Kohn anomalies and has directly been observed using angular
correlation of electron-positron annihilation radiation in
YNi$_2$B$_2$C and LuNi$_2$B$_2$C\cite{Bullock98, Zarestky99,
Dugdale99}. Moreover, the antiferromagnetic ordering found in the
borocarbides with a magnetic rare earth has a wavevector very
close to the nesting vector Q$\approx(0.5,0,0)$ \cite{Lynn, Rhee}
(in R=Er, Ho and Dy; in Tm it appears in a magnetic field phase).
Nesting is generally considered to play against the formation of a
phonon mediated superconducting state, as evidenced in other
families of superconductors (e.g. in the
dichalchogenides\cite{Nalseiro, CastroNeto}). Here we present
measurements of tunneling spectroscopy of the non-magnetic
YNi$_2$B$_2$C and LuNi$_2$B$_2$C, which show that the highly
anisotropic electron-phonon interaction produced by Fermi surface
nesting drives the system to an also anisotropic superconducting
state. We give new insight into the nature of the pairing
mechanism, the order parameter symmetry and the gap anisotropy.

Tunneling spectroscopy is in principle one of the most powerful
experimental tools to investigate the anisotropy of the
superconducting gap and to obtain information about the pairing
mechanism\cite{Parks}. We use the same scanning tunneling
microscopy and spectroscopy (STM/STS) set-up as in
ref.\cite{Suderow01}, where we studied the magnetic
TmNi$_2$B$_2$C, with an improved resolution allowing now
measurements down to 0.5K. We have also characterized other
materials with the same set-up (Al\cite{Suderow02a}, Pb and
NbSe$_2$) and demonstrated that this is indeed the actual
temperature of tip and sample and that we do not need any
additional pair breaking parameter $\Gamma$\cite{Yokoya00} to
explain our data. Previous works about the spectroscopy of
borocarbides had a lower resolution in energy\cite{Yokoya00,
Ekino, Wilde, Sakata}. The sample is broken in air on the sample
holder of the STM and cooled down immediately. The resulting
surface presents the same topology as for the Tm borocarbide,
which consists of inclined planes and bumps, typical of a
conchoidal fracture, presenting no clear crystallographic
orientation\cite{Suderow01}. So it is crucial to characterize the
superconducting behavior in well-differentiated regions of the
sample. We use a home made x-y table that gives the possibility to
change in-situ the scanning window (of 1$\times$1$\mu$m$^2$) in an
area of 2$\times$2mm$^2$. We measured three different samples (in
three different cool-downs) of each compound. In the case of
LuNi$_2$B$_2$C, all of them were freshly broken pieces of the same
single crystal grown by a flux technique described in
ref.\cite{Canfieldgeneral}. The YNi$_2$B$_2$C samples came from
two different single crystals, one grown by the same flux
technique and the other in an image furnace. The tunneling
conditions were always very good, with high valued measured work
functions of several eV.

\begin{figure}[ht]
\includegraphics[width=8cm,clip]{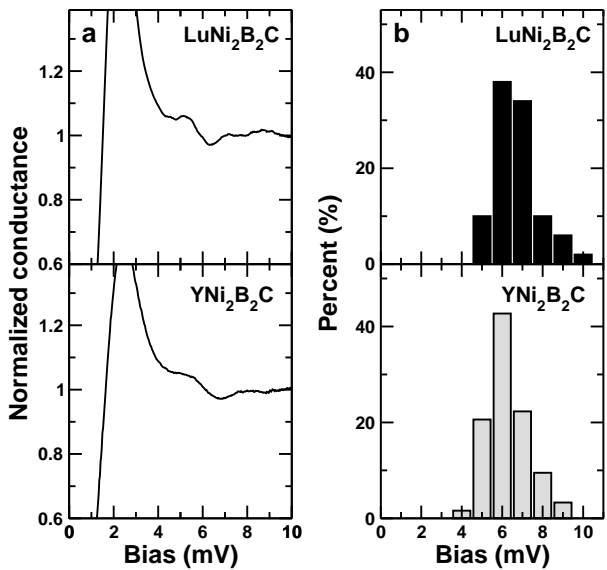}
\caption{
a) Tunneling conductance spectra (taken at 0.5K and normalized to
the conductance value at high bias voltage) where electron-phonon
coupling features at high voltages clearly appear. b) Histogram
counting the number of times we find such a feature (within a
sample of a hundred spectra), with the maximum value of the
derivative of the conductance at the voltage reported on the x
axis.}
\label{fig:Fig1}
\end{figure}

The tunneling differential conductance, $\sigma$=dI/dV, between a
normal metal and a superconductor gives a direct (temperature
smeared) measurement of the superconducting density of states.
Therefore, data show the eventual presence of very low energy
excitations within the superconducting gap, $2\Delta$, and a high
quasiparticle peak at voltages close to the gap. But if the
measurement is sufficiently precise, one can also try to resolve
tiny features at voltages corresponding to the sum of $\Delta$ and
the characteristic energy of the bosonic excitations leading to
superconductivity, in order to obtain information about the nature
of these excitations\cite{Parks, Carbotte90, Suderow02b}. We could
indeed observe these features in the tunneling conductance spectra
for the Y and Lu borocarbides. In Fig.1 we show a typical curve
together with a histogram where the number of times that such
features are observed is presented as a function of the bias
voltage (their voltage position is determined from the maximum of
the derivative of the conductance above the gap\cite{Parks,
Carbotte90, Suderow02b}). Most of them appear centered at a
voltage between 6 and 7mV. Substracting the values of $\Delta$, as
estimated below, the maxima become centered between 4 and 5 mV.
This is precisely the energy at which a high peak develops in
neutron scattering experiments due to a low energy phonon
mode\cite{Bullock98, Zarestky99}. An estimation of the electron
phonon coupling constant $\lambda$ from our data gives rather
high values between 0.5 and 0.8 in both compounds\cite{note}. On the other
hand, previous estimations using thermodynamic measurements, which
take into account the Eliashberg function at all relevant phonon
frequencies, but which do not directly specify the nature or
spectral weight of each mode, have given values of $\lambda$
between 0.75 and 1.2\cite{Manalo,Michor}. The fact that we find
values of $\lambda$ smaller but very close to the ones estimated
with thermodynamic measurements means that the structures in the
Eliashberg function associated with higher energy phonon modes
\cite{El-Hagary,Shulga} are more spread-out in energy than the
peak corresponding to the 4mV phonon mode and fall below our
experimental resolution. It also means that the low energy phonon
mode measured here is essential in the formation of
superconducting correlations.

The most striking point however is that the phononic density of
states at these energies results from a mode having a wavevector
comparable to the nesting vector Q (0.5,0,0) \cite{Rhee,
Dugdale99}. Neutron scattering experiments show indeed pronounced
Kohn anomalies\cite{Bullock98, Zarestky99}, where the nesting
feature of the Fermi surface\cite{Rhee, Dugdale99} leads to a
significant softening, when decreasing temperature, of the
low-lying transverse phonon branches at wavevectors close to
(0.5,0,0). This behavior results in a strong and highly
anisotropic electron-phonon coupling. Our experiment demonstrates
for the first time that this anisotropic electron-phonon
interaction, produced by Fermi surface nesting, leads to
superconducting correlations, having an important contribution to
the total electron-phonon coupling constant $\lambda$.
Correspondingly, we can expect the superconducting gap to be also
highly anisotropic.

\begin{figure}[ht]
\includegraphics[width=8cm,clip]{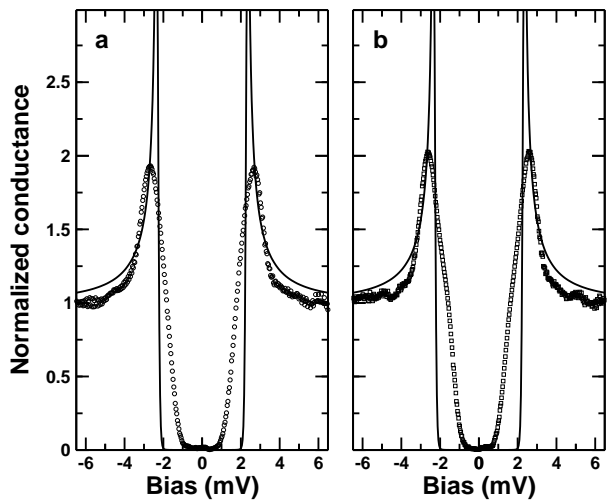}
\caption{ Tunneling conductance spectra in the Lu (a) and Y (b)
borocarbides at 0.5K normalized to the conductance value at high
bias voltage (tunneling resistance of 2M$\Omega$). The full lines
are fits to conventional BCS theory ($\Delta$=2.3meV).}
\label{fig:Fig2}
\end{figure}

Tunneling conductance measurements done with a STM give the
superconducting density of states averaged over part (but not all)
of the Fermi surface, depending on the relative position of the
tip onto the sample\cite{Chen}. In most cases we measure tunneling
conductance curves as the ones shown in Fig.2, where the
conductance is zero below 0.8mV and then increases up to a peak
located at 2.6mV in Lu and 2.3mV in Y. These curves cannot be
fitted by conventional BCS theory. The discrepancy with most
simple BCS theory is not due to a lifetime smeared BCS density of
states, which leads to a non zero density of states at low
energies, not observed in our data. The situation is clearly very
different from the one found in the very similar borocarbide
material TmNi$_2$B$_2$C \cite{Suderow01}, where the spectra can be
fitted by BCS theory. The form of the curves in Fig.2 shows that
the electrons contributing to the tunneling current at this
precise location come from parts of the Fermi surface with a
continuous distribution of values of the superconducting
gap\cite{Hess90,Hayashi}. This reveals that the gap function must
be anisotropic. Moreover, the spectra in (Y,Lu)Ni$_2$B$_2$C remain
with the same shape over regions much larger than the coherence
length (usually dimensions about 200$\times$200nm$^2$ or bigger
whereas $\xi$$\sim$7nm \cite{Izawa01}), maintaining
their overall form when we increase the temperature, and becoming
completely flat at the bulk critical temperature. Similar
observations have been made in the anisotropic superconductor
NbSe$_2$ \cite{Hess90,Hayashi}, where the spectra have,
qualitatively, the same form.

On the other hand, although the behavior shown in Fig.2 appears
most frequently on the surface, we can also find different
topographical regions of similar sizes with very different
spectra, as shown in Fig.3a. In Fig.3b we compare the temperature
dependence of $\Delta$, as estimated from the voltage position of
the maximum of d$\sigma$/dV, and in Fig.3c the corresponding
tunneling density of states at 0.5K in three different locations
of YNi$_2$B$_2$C (the same is observed in LuNi$_2$B$_2$C).
Clearly, the differences in the spectra are associated with
smaller values of the superconducting gap and also a reduced
critical temperature. Remarkably, the mean value of the
superconducting gaps shown in Fig.3b follows well the temperature
dependence predicted by simple BCS theory (lines). What is more,
the curves corresponding to a smaller critical temperature in
Fig.3c (top curve) approach much better an isotropic BCS {\it
s}-wave behavior than those corresponding to the bulk T$_c$
(bottom curve). This must be related to changes in the anisotropy
as a function of the local T$_c$. To obtain more quantitative
information about this effect we have fitted the experiment to a
modified BCS {\it s}-wave density of states assuming that the
dispersion in the values of the superconducting gap can be
modelled by a gaussian distribution centered around $\Delta_0$
with a width of $\varepsilon$ (a similar approach has been
previously used in the anisotropic superconductor
NbSe$_2$\cite{Hayashi}). The dashed lines in Fig.3c give the best
fit to this model. The agreement with the experiment is much
better in regions with a smaller T$_c$, where $\varepsilon$
decreases. The inset in Fig.3b represents the dependence of the
estimated anisotropy, $\varepsilon/\Delta_0$, as a function of the
measured critical temperature in several topographical regions of
the same samples. Note that these values do not give an indication
of the whole distribution of values of the superconducting gap, as
a given STM spectrum is a local measurement which probes only part
of the Fermi surface\cite{Chen}. Therefore, our data are
in good agreement to previous macroscopic measurements
\cite{Nohara00, Izawa01, Boaknin01, Izawa02, Jacobs, Yokoya00,
SangYang, Cheon, Gammel}.

\begin{figure}[ht]
\includegraphics[width=8cm,clip]{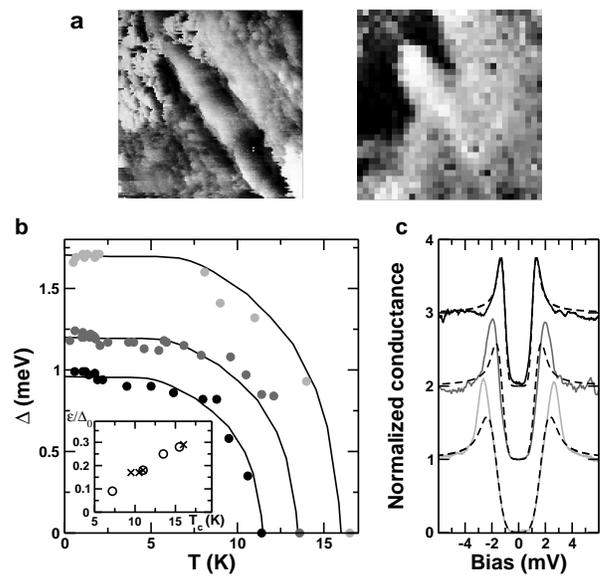}
\caption{ a) Topografic (left) and STS (right) images at 0.5K
(YNi$_2$B$_2$C, 310$\times$310nm$^2$). The latter figure is
indicative of the superconducting density of states as a function
of the position of the tip on the surface\cite{Sakata, Suderow01}.
The contrast is adjusted such that bright spots correspond to
$\sigma$(V) similar to Fig.2, and dark spots to $\sigma$(V)
similar to the top curve in Fig.3c. b) Temperature dependence of
the mean value of the superconducting gap (points) in
YNi$_2$B$_2$C. We take $\Delta$ as the voltage where
d$\sigma$(V)/dV is maximum. Lines give the temperature dependence
of the superconducting gap expected in BCS theory. The inset
represent the variation of $\varepsilon/\Delta_0$ (see text) with
T$_c$ for YNi$_2$B$_2$C (circles) and LuNi$_2$B$_2$C (crosses). c)
Spectra taken at 0.5K for each temperature run (same color for the
straight lines as points in b), shifted for clarity. Dashed lines
are fits to the BCS theory with anisotropy (see text), where, from
top to bottom, $\Delta_0$=1.1mV and $\varepsilon$=0.2mV,
$\Delta_0$=1.3mV and $\varepsilon$=0.32mV and $\Delta_0$=1.8mV and
$\varepsilon$=0.5mV.} \label{fig:Fig3}
\end{figure}

Anisotropic superconductivity is expected to be very sensitive to
even non-magnetic defects that reduce T$_c$ and decrease the
anisotropy\cite{Carbotte81,Golubov97}. Therefore, in an
anisotropic superconductor, local measurements in topographically
different positions on the surface can in principle show different
forms of the superconducting gap and values of the local critical
temperature, associated with the presence of defects lying near
the surface or with an irregular topography\cite{Bascones02}.
Isotropic superconductors, by contrast, present BCS spectra and
the bulk critical temperature over the whole
surface\cite{Suderow01,Suderow02a,Pan}.

In (Y,Lu)Ni$_2$B$_2$C we observe a gradual decrease of the local
critical temperature down to about half its bulk value {\it and}
of the anisotropy (inset of Fig.3b) in different locations which
also show a different topography (Fig.3a). Our measurements show a
direct correlation between the local depression of T$_c$ with a
local change of the form of the gap. This also demonstrates
that the superconducting gap must be highly anisotropic, and it
provides an additional test of the nature of the order parameter
in these materials. If the anisotropy is due to a {\it d}-wave
order parameter, the defects tend to suppress superconductivity
altogether and low energy excitations appear, filling the density
of states below the gap. If, however, the superconducting wave
function is anisotropic {\it s}-wave, the defects tend to suppress
the anisotropy, leading to a more isotropic gap and a decreased
critical temperature\cite{Hirschfeld, Norman}. The curves plotted
in Fig.3c, with no low energy excitations, definitely imply that
superconductivity in YNi$_2$B$_2$C and LuNi$_2$B$_2$C is highly
anisotropic but {\it s}-wave.

Using the same experimental protocol, we find a completely
different behavior in the chemically very similar magnetic
TmNi$_2$B$_2$C compound (T$_c$=11K), where BCS {\it s}-wave like
spectra without a significant anisotropy (and bulk critical
temperature) are measured over the whole surface\cite{Suderow01}.
However, we can state that the surface is analogous in all three
compounds, because the measured work function and the
topographical images are similar. This could be interpreted as an
indication that mechanisms inducing anisotropy in YNi$_2$B$_2$C
and LuNi$_2$B$_2$C are not operating in this borocarbide, or that
the intrinsic magnetic disorder already smears out homogeneously
the anisotropy.

In conclusion, we have studied the tunneling spectroscopy in the
non-magnetic borocarbides using high resolution STM/STS. We have
been able to characterize important microscopic aspects of the
superconducting state, which is an anisotropic {\it s}-wave state,
where a significant part of the electron-phonon coupling leading
to superconducting correlations is also highly anisotropic and due
to soft phonons. The demonstration of this new mechanism
strengthens the hope for further discoveries in the area of new
high-T$_c$ superconducting materials.

We acknowledge discussions with F. Guinea, A. I. Buzdin, J.
Flouquet, A. Levanyuk and support from the ESF programme VORTEX,
from the MCyT (Spain; grant MAT-2001-1281-C02-0), from FERLIN and
from the Comunidad Aut\'onoma de Madrid (Spain). The Laboratorio
de Bajas Temperaturas is associated to the ICMM of the CSIC. Ames
Laboratory is operated for the U. S. Department of Energy by Iowa
State University under Contract No. W-7405-Eng-82. This work was
supported by the Director for Energy Research, Office of Basic
Energy Sciences.

\end{document}